\documentclass{article}
\usepackage{spconf,amsmath,amssymb,graphicx}

\usepackage{enumitem}
\setlist{nosep, leftmargin=14pt}
\usepackage{amsmath}
\usepackage{xcolor}
\usepackage[normalem]{ulem}
\usepackage{mwe} 
\usepackage{graphicx}

\title{Clinical Interpretability of Deep Learning Segmentation Through Shapley-Derived Agreement and Uncertainty Metrics}

\name{Tianyi Ren$^{1*}$, Daniel Low$^{2*}$, Pittra Jaengprajak$^{2}$, Juampablo Heras Rivera$^{1}$, Jacob Ruzevick$^{3}$, Mehmet Kurt$^{1}$}
\address{$^{1}$Department of Mechanical Engineering, University of Washington, Seattle, WA\\
$^{2}$University of Washington School of Medicine, Seattle, WA\\
$^{3}$Department of Neurological Surgery, University of Washington, Seattle, WA\\
$^{*}$Equal contribution}
\begin{document}
%
\maketitle

\vspace{0.43in}

\begin{abstract}
Segmentation is the identification of anatomical regions of interest, such as organs, tissue, and lesions, serving as a fundamental task in computer-aided diagnosis in medical imaging. Although deep learning models have achieved remarkable performance in medical image segmentation, the need for explainability remains critical for ensuring their acceptance and integration in clinical practice, despite the growing research attention in this area. Our approach explored the use of contrast-level Shapley values, a systematic perturbation of model inputs to assess feature importance. While other studies have investigated gradient-based techniques through identifying influential regions in imaging inputs, Shapley values offer a broader, clinically aligned approach, explaining how model performance is fairly attributed to certain imaging contrasts over others. Using the BraTS 2024 dataset, we generated rankings for Shapley values for four MRI contrasts across four model architectures. Two metrics were proposed from the Shapley ranking: agreement between model and ``clinician" imaging ranking, and uncertainty quantified through Shapley ranking variance across cross-validation folds. Higher-performing cases (Dice \textgreater0.6) showed significantly greater agreement with clinical rankings. Increased Shapley ranking variance correlated with decreased performance (U-Net: $r=-0.581$). These metrics provide clinically interpretable proxies for model reliability, helping clinicians better understand state-of-the-art segmentation models.

\end{abstract}


\section{Introduction}
\label{sec:intro}

\begin{figure}[htbp]
\centering
\includegraphics[width=\columnwidth]{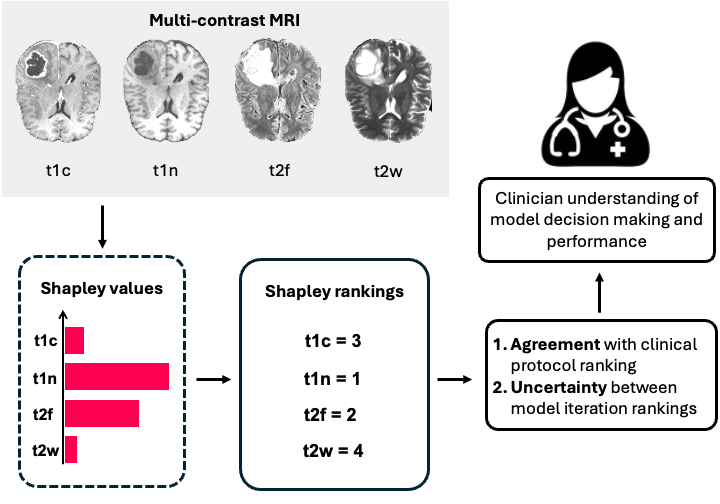}
\caption{Overview of explainability metrics derived from model generated Shapley values.}
\label{}
\end{figure}

Computer-aided diagnosis relies on segmentation, an essential task that involves the precise delineation of anatomical structures \cite{hesamian2019deep}. For more complex pathologies, such as gliomas, segmentation is typically carried out using multi-contrast MRI imaging to visualize unique structural features \cite{10980803}. Despite advanced capabilities in segmentation tasks for complex pathologies, many deep learning models are considered ``black boxes" with limited insight into decision making, especially from a clinical perspective \cite{rudin2019stop}. Previous studies have explored the use of gradient-based techniques, such as Grad-CAM, to visualize influential regions for predictions \cite{selvaraju2017grad}. While useful, these methods require thresholding decisions to identify important pixels, which can significantly affect interpretation. In an effort to emulate the algorithm used by radiologists to identify tumor regions through MRI contrast differences, we generated contrast-level Shapley values to serve as explanations of model segmentation \cite{DBLP:conf/xai/RenROPCRK25}. This method fairly attributes imaging contrasts through the systematic perturbation of model inputs and assessment of impact on model performance, assigning quantitative values to each MRI contrast type.

To bridge the gap between Shapley values and clinical decision making, we investigated the considerations of ``agreement" and  ``uncertainty" that influence the identification and diagnosis of lesions. Agreement refers to the consensus between clinicians and standard imaging protocols, while uncertainty is the variation between clinician opinions \cite{carney2004radiologist}. We mimic these factors within models by converting Shapley values to a ranking that answers the question: Which MRI contrast provides the most value in segmenting regions of brain tumors? Agreement was then assessed through accordance between model and clinical based rankings of MRI contrast, while uncertainty was evaluated through the variance of ranking between model data folds. Our contributions are as follows:
\begin{enumerate}
  \item The development of an agreement metric for assessing model explanation alignment with clinical protocols.
  \item The assessment of model uncertainty that mirrors interclinician disagreement.
\end{enumerate}
These benchmarks extrapolate the explainative value of contrast-level Shapley values, allowing clinicians to understand if a model of interest aligns with clinical protocols and can provide consistency in segmentation predictions. 

\section{Methods}
\label{sec:format}
\textbf{2.1 Dataset and Model Architecture}\vspace{0.5em}

\noindent Brain tumor segmentation performance was assessed on the Brain Tumor Segmentation (BraTS) Challenge 2024 GoAT challenge, consisting of 1351 subjects \cite{de20242024}. Each subject included four MRI contrasts: T1c, T1n, T2f, and T2w. Ground truth annotations included three tumor sub-regions: necrotic core, edema, and enhancing tumor segmentation and performance was evaluated using Dice similarity coefficient. Four deep learning architectures were then evaluated using five-fold cross-validation: U-Net, Seg-Resnet, UNETR, and Swin-UNETR \cite{DBLP:conf/xai/RenROPCRK25,ren2023optimization,ren2024re}. \vspace{0.5em}

\noindent\textbf{2.2 Contrast Level Shapley Value}\vspace{0.5em}

\noindent The contrast-level Shapley value framework is used to provide intuitive, quantitative explanations for our multi-contrast segmentation models \cite{DBLP:conf/xai/RenROPCRK25}. Given the four 3D-MRI ($N$) contrasts as a multi-channel input $I \in \mathbb{R}^{N \times D \times W \times H}$, the deep learning models ($\omega$) are trained to predict the tumor labels $\hat{x}_0$ where $\hat{x}_0 = \omega(I)$.

The contrast level Shapley value $\phi_i(D)$ quantifies the contribution of a specific contrast ($i$) to the Dice score ($D$)  by weighting the contributions over all possible subsets ($S$) of contrasts:

\begin{equation}
\phi_i(D) = \sum_{S \subseteq N \setminus n_i} \frac{|S|!(|N| - |S| - 1)!}{|N|!} \left(D(S \cup n_i) - D(S)\right)
\label{eq:shapley_value}
\end{equation}

Let $N = \{n_1, \dots, n_{|N|}\}$ be the set of contrasts, with cardinality $|N|$. Let $n_i \in N$ denote the $i$-th contrast, where $i \in \{1, \dots, |N|\}$. We define $D(S)$ as the target metric evaluated on any subset $S \subseteq N$. These quantitative $\phi_i(D)$ values were produced for the tumor sub-region predictions for each imaging contrast, across all folds and model architectures. Overall tumor Shapley values were derived from an average of the sub-region Shapley values.\vspace{0.5em}

\noindent\textbf{2.3 Clinical Imaging and Model Ranking Agreement}\vspace{0.5em}

\noindent The BraTS 2021 Brain Tumor Segmentation Benchmark paper is used as the protocol for clinical reasoning and decision making for human ranking of the BraTS Challenge 2024 GoAT challenge dataset \cite{baid2021rsna}. We randomly selected 61 cases from the U-Net model's ``low" Dice score group (Dice below 0.5), which was half of that group. We also selected 40 cases from the ``high" Dice score group (Dice above 0.5).

From this cohort of cases, two U.S. medical students independently followed the Brats 2021 protocol and assigned rankings of importance for the four MRI contrasts with regards to tumor region identification. Each imaging type was assigned a value from 1-4 (1 indicating highest importance) with ties being possible. Consensus rankings were created by averaging the ranks from both annotators for each contrast and applying dense ranking. Furthermore, as expressed explicitly in the protocol, a general clinical ranking was assigned to the same cases to serve as a “clinical standard” imaging ranking (T1c=1, T2f=2, T1n/T2w=3) \cite{baid2021rsna}. Then, for each subject in the cohort, overall tumor Shapley values for each imaging was averaged over the five folds and converted to a Shapley ranking ($R_{\phi_i(D)}$) from 1-4 (1 indicating greatest raw contrast level Shapley value).\vspace{0.5em}

\noindent\underline{Normalized Spearman Footrule Distance (NSF)}: 
Agreement for each subject was quantified using Normalized Spearman Footrule  \cite{diaconis1977spearman}:
\begin{equation}
\textit{NSF} = 1 - (1/D_{\text{max}}) \sum_{i=1}^{|N|} |R_{\phi_i(D)} - R_{\kappa_i}|
\end{equation}
where $|N|$  = 4 MRI contrasts, $R_{\phi_i(D)}$ and $R_{\kappa_i}$ are the rankings (1-4) assigned to the $i$-th contrast by the model and student consensus or clinical standard, respectively, and $D_{\text{max}}$ = 8 is the maximum possible sum of rank differences. This metric ranges from 0 (complete disagreement) to 1 (perfect agreement). Agreement was calculated for each subject and correlated with average Dice score performance across folds. The Mann-Whitney U test \cite{mcknight2010mann}, a ranked-comparison analysis, was performed to compare agreement between the Dice groups above 0.5 and the $<0.5$ Dice group ($p < 0.05$).\vspace{0.5em}



\noindent\textbf{2.4 Shapley Ranking Interfold Variance Calculation}\vspace{0.5em}

\noindent Let $r_{\Phi_i(D), k}$ denote the rank (1-4) of the tumor sub-region Shapley value for the $i$-th contrast within the $k$-th fold, where $i \in \{1, \dots, |N|\}$ and $N=4$ and $k \in \{1, \dots, K\}$ and $K=5$ (see 2.3). Sub-region variance ($v$) across a model's five folds and four contrasts was then calculated for each subject:
\begin{equation}
\textit v = (1/|N|) \sum_{i=1}^{|N|} \text{Var}(R_i)
\end{equation}
where $|N|$ = 4 MRI contrasts. For each MRI contrast $i$, $R_i$ represents the vector of 5 ranks ($K$, one per fold), and Var($R_i$) is the variance of these ranks across folds, calculated using the standard variance formula:
\begin{equation}
\text{Var}(R_i) = 1/(K-1) \sum_{k=1}^{K} (r_{\Phi_i(D), k} - \bar{r}_i)^2
\end{equation}
where $K$ = 5 folds and and $\bar{r}_i$ is the mean rank of contrast $i$ across all folds. $v$ was calculated for each tumor sub-region and averaged to obtain an overall variance ($V${= $\bar{v}$}) for Shapley rankings. Overall, low variance indicates consistent contrast weighing across folds, while high variance suggests inconsistent weighing. \vspace{0.5em}

\noindent\underline{Bootstrap Analysis}: Shapley rank variance ($V$) distributions across all models were positively skewed, with the majority of subjects clustered at low variance values. To address this class imbalance when assessing the relationship between variance and Dice scores, we employed bootstrap resampling \cite{bland2015statistics} with replacement of the low variance group (80th percentile). 

We performed 5000 bootstrap iterations. In each iteration, subjects were randomly sampled with replacement to create datasets where each sample maintained the original variance distribution. Spearman correlations \cite{wissler1905spearman} between Shapley rank variance and average Dice scores across folds were calculated for each bootstrap sample, yielding a distribution of correlation coefficients. To obtain greater interpretation from a given Shapley rank variance value, we also performed a subcohort Spearman correlation ($p < 0.05$) thresholded at a variance of 0.275 (this value was chosen to provide balanced groups between all four models). 

\begin{figure}[t]
\centering
\includegraphics[width=\columnwidth]{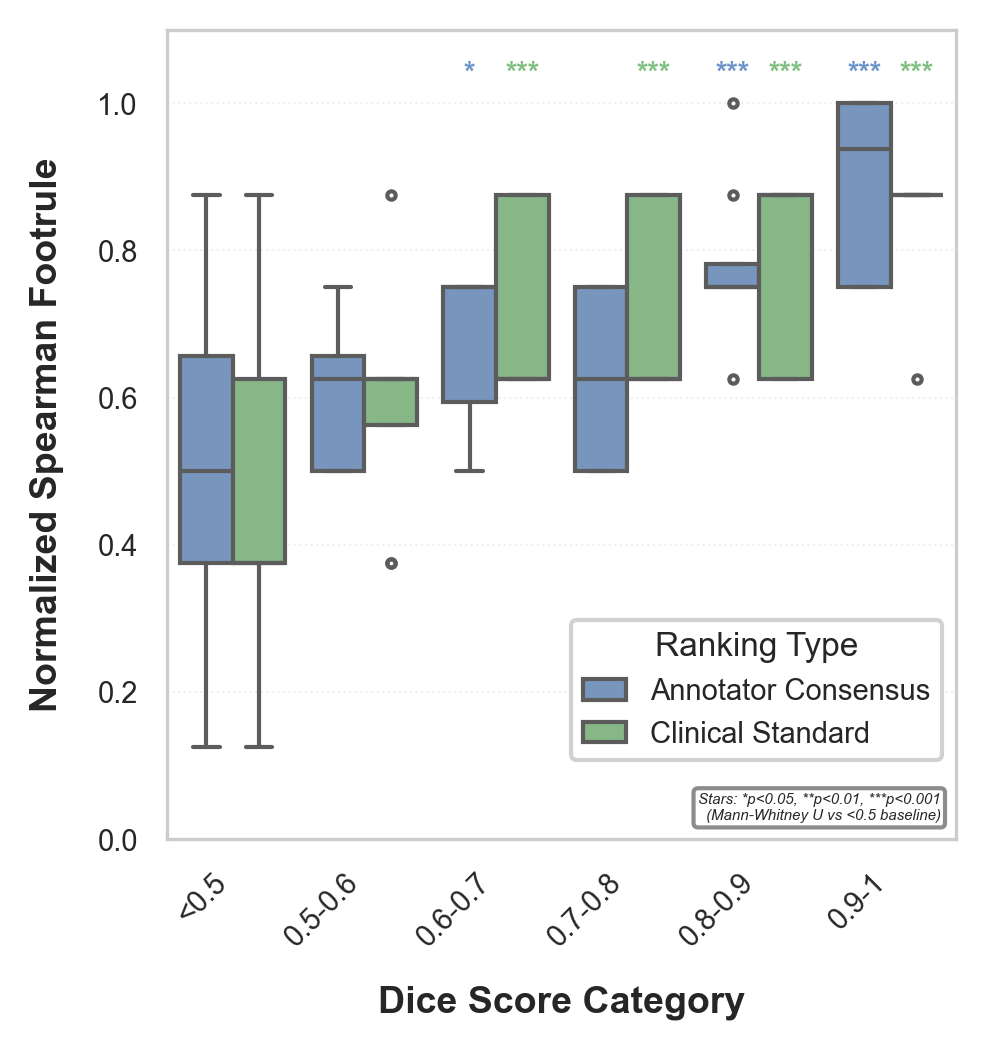}
\caption{Comparison of Normalized Spearman Footrule (agreement) between increasing Dice score groups. Colors indicate agreement between U-Net Shapley ranking and annotator consensus (blue) and clinical standard (green)}
\label{fig:unet_agreement}
\end{figure}

\section{Results}
\label{sec:results}

Subjects with higher average Dice score across folds had greater agreement between U-Net and annotator and clinical standard image ranking (Figure 2) using the Mann-Whitney U test. Specifically, the 0.6-0.7 ($p<0.05$), 0.8-0.9 ($p<0.001$), and 0.9-1 ($p<0.001$) Dice groups had significantly greater agreement than the $<0.5$ Dice group for the annotator consensus ranking. All groups above 0.6 Dice had significantly greater agreement than the $<0.5$ Dice group for the clinical standard ranking ($p<0.001$). 

Increases in Shapley ranking variance demonstrated decreases in average Dice score across all models (Figure 3). Bootstrap correlations calculated at the 80th percentile Shapley rank variance of the cohort for U-Net, Seg-Resnet, Swin-UNETR, and UNETR models were -0.581, -0.604, -0.351 and -0.351, respectively ($p<0.05$). 95\% confidence intervals are shown in the figure by the shaded grey expansion of the correlation vector. While all model types show correlations, they are stronger in U-Net and Seg-Resnet compared to Swin-UNETR and UNETR. Inter-cohort Spearman correlation of each model was also performed stratified at 0.275 Shapley ranking variance, resulting in the following r and p-value pairs for each model ($<0.275$ group; $>0.275$  group): U-Net ($ r = 0.167$, $p <0.001$; $r = -0.538$, $p < 0.001$), Seg-Resnet ($r = 0.083$, $p < 0.05$; $r = -0.398$, $p < 0.001$), Swin-UNETR ($r = 0.067$, $p > 0.05$; $r = 0.251$, $p < 0.001$), and UNETR ($r = 0.165$, $p < 0.001$; $r = -0.244$, $p < 0.001$).

\begin{figure*}[t]
\centering
\includegraphics[width=0.8\textwidth]{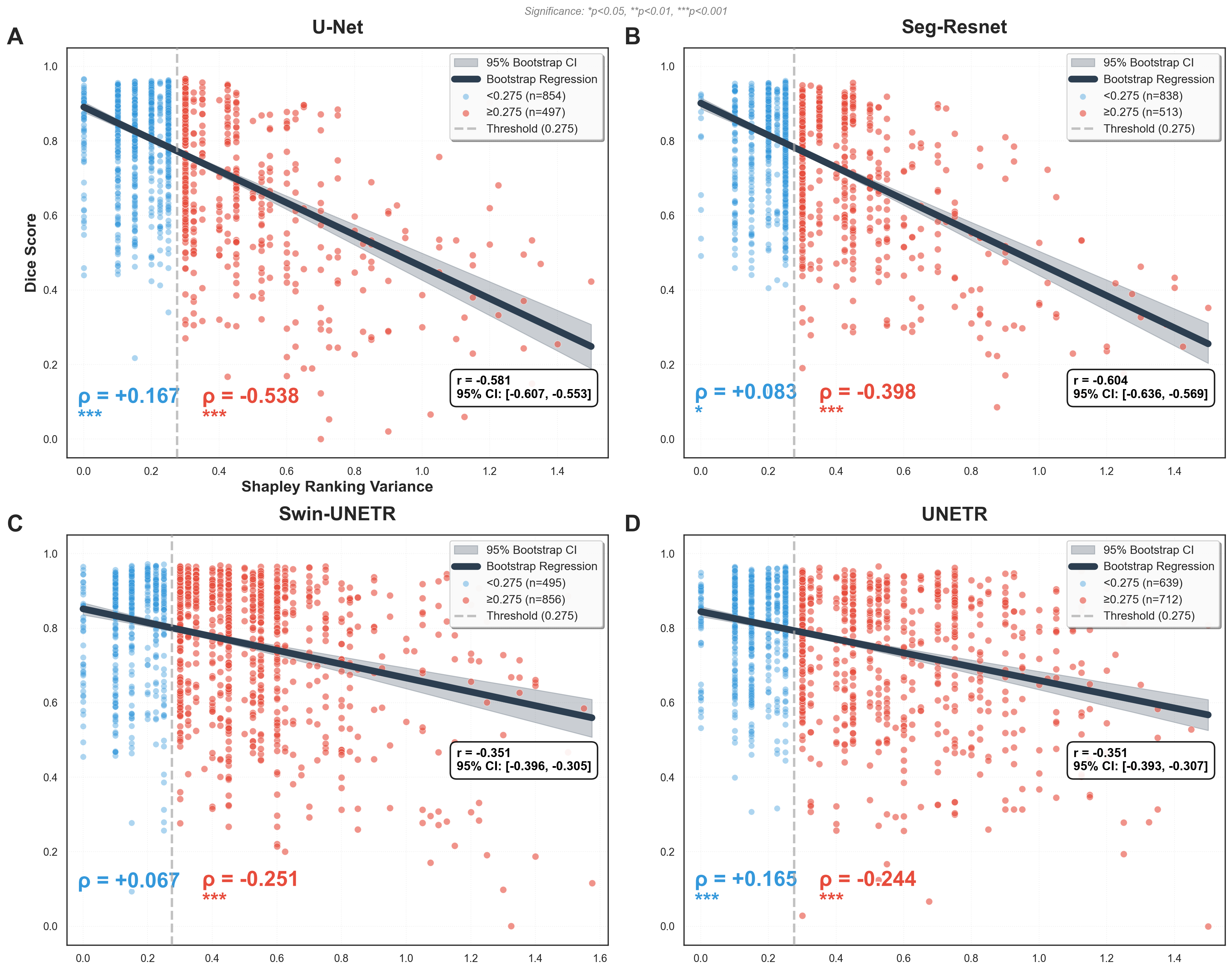}
\caption{Dice score performance correlated against Shapley rank variance ($V$, see 2.4) for four models. Subcohort Spearman regression values listed above (red) and below (blue) threshold at 0.275 Shapley rank variance. }
\label{fig:bootstrap_variance}
\end{figure*}

\section{Discussion}
\label{sec:typestyle}

\noindent The use of both annotator consensus and clinical standard rankings was to capture the concept of a diagnostic algorithm that radiologists utilize when identifying pathologies. In this analysis, our goal is to simulate the agreement between the model's emphasis on different imaging contrast and that with which a radiologist would use. The results demonstrate that subjects with higher Dice scores had significantly greater agreement between their Shapley and respective ranking compared to the $<0.5$ Dice group. In practical terms, the agreement between Shapley and a radiologist's general ranking protocol (provided to the model) could provide a clinically oriented explanation of the model's performance, allowing the clinician to understand the reasoning and subsequent prediction.

Variance of Shapley rank across folds served as a metric for assessing model uncertainty for each case. As there is always a degree of uncertainty inherent to any subject (e.g. radiologists having differing opinions), we considered this analysis a valid examination of model behavior \cite{carney2004radiologist}. In addition to finding greater uncertainty correlating with Dice score degradation, we calculated correlations of Dice and uncertainty in two bins thresholded at 0.275 Shapley rank variance to extract greater interpretive value from a given uncertainty. This analysis yielded weakly positive or no correlations within the $<0.275$ group, but significant negative correlations in the $>0.275$ group, reflecting the greater utility of a high versus a low uncertainty value. If the model's uncertainty is low, the insight into Dice value is not meaningful as Dice values are relatively stable in this group. Conversely, if the model's uncertainty is high, we understand that the model performance is likely poor.

A limitation of this study is the lack of an agreement consensus sourced from radiologists. Although we provided two means of comparison (medical student annotator consensus and paper-derived clinical standard) for the Shapley ranking agreement analysis, we aim to investigate agreement through more robust clinical comparisons. In addition, the idea of uncertainty in model decisions is subjective and can be approached from many angles. Future directions include the use of Monte Carlo dropout to quantify prediction uncertainty across multiple stochastic forward passes. This analysis performed outside of Shapley rankings could potentially allow prediction of model performance outside of training settings. 

\section{Conclusion}

This study demonstrates that Shapley-derived explanations can serve as indicators of automated segmentation performance. Agreement between model contrast prioritization and clinical protocols correlates with better performance and greater model uncertainty points to performance degradation. These Shapley-derived explanations thus help bridge the disaparity between opaque model reasoning and clinically-aligned explanations, building greater confidence in the utility and performance of segmentation models. 

\section{Compliance with ethical standards}
This research study was conducted using de-identified, publicly available 
data from the Brain Tumor Segmentation (BraTS) 2024 GoAT Challenge dataset, 
accessed through the Synapse platform. The dataset is distributed under a 
Creative Commons Attribution license for research purposes. Ethical approval 
was not required for this retrospective analysis of de-identified publicly 
available data, as confirmed by the data use agreement.
\section{Acknowledgements}
This research is funded by NSF Grant No.CMMI-1953323. 

\label{sec:page}




\bibliographystyle{IEEEbib}
\bibliography{strings,refs}

\begin{thebibliography}{10}

\bibitem{hesamian2019deep}
M.~H. Hesamian, W.~Jia, X.~He, and P.~Kennedy,
\newblock ``Deep learning techniques for medical image segmentation:
  achievements and challenges,''
\newblock {\em Journal of digital imaging}, vol. 32, no. 4, pp. 582--596, 2019.

\bibitem{10980803}
L.~Yan, C.~Wang, F.~Zhong, and Y.~Wang,
\newblock ``Clinical inspired mri lesion segmentation,''
\newblock in {\em 2025 IEEE 22nd International Symposium on Biomedical Imaging
  (ISBI)}, 2025, pp. 1--4.

\bibitem{rudin2019stop}
C.~Rudin,
\newblock ``Stop explaining black box machine learning models for high stakes
  decisions and use interpretable models instead,''
\newblock {\em Nature machine intelligence}, vol. 1, no. 5, pp. 206--215, 2019.

\bibitem{selvaraju2017grad}
R.~R. Selvaraju, M.~Cogswell, A.~Das, R.~Vedantam, D.~Parikh, and D.~Batra,
\newblock ``Grad-cam: Visual explanations from deep networks via gradient-based
  localization,''
\newblock in {\em Proceedings of the IEEE international conference on computer
  vision}, 2017, pp. 618--626.

\bibitem{DBLP:conf/xai/RenROPCRK25}
T.~Ren, J.~H. Rivera, H.~Oswal, Y.~Pan, A.~Chopra, J.~Ruzevick, and M.~Kurt,
\newblock ``Here comes the explanation: A shapley perspective on multi-contrast
  medical image segmentation,''
\newblock in {\em Joint Proceedings of the xAI 2025 Late-breaking Work, Demos
  and Doctoral Consortium co-located with the 3rd World Conference on
  eXplainable Artificial Intelligence (xAI 2025), Istanbul, Turkey, July 9-11,
  2025}, ser. CEUR Workshop Proceedings, vol. 4017, pp. 185--192, CEUR-WS.org,
  2025.

\bibitem{carney2004radiologist}
P.~A. Carney, J.~G. Elmore, L.~A. Abraham, M.~S. Gerrity, R.~E. Hendrick, S.~H.
  Taplin, W.~E. Barlow, G.~R. Cutter, S.~P. Poplack, and C.~J. D'Orsi,
\newblock ``Radiologist uncertainty and the interpretation of screening,''
\newblock {\em Medical decision making}, vol. 24, no. 3, pp. 255--264, 2004.

\bibitem{de20242024}
M.~C. de~Verdier, R.~Saluja, L.~Gagnon, D.~LaBella, U.~Baid, N.~H. Tahon,
  M.~Foltyn-Dumitru, J.~Zhang, M.~Alafif, S.~Baig, et~al.,
\newblock ``The 2024 brain tumor segmentation (brats) challenge: Glioma
  segmentation on post-treatment mri,''
\newblock {\em arXiv preprint arXiv:2405.18368}, 2024.

\bibitem{baid2021rsna}
U.~Baid, S.~Ghodasara, S.~Mohan, M.~Bilello, E.~Calabrese, E.~Colak,
  K.~Farahani, J.~Kalpathy-Cramer, F.~C. Kitamura, S.~Pati, et~al.,
\newblock ``The rsna-asnr-miccai brats 2021 benchmark on brain tumor
  segmentation and radiogenomic classification,''
\newblock {\em arXiv preprint arXiv:2107.02314}, 2021.

\bibitem{ren2024re}
T.~Ren, A.~Sharma, J.~H. Rivera, H.~Rebala, E.~Honey, A.~Chopra, J.~Ruzevick,
  and M.~Kurt,
\newblock ``Re-diffinet: modeling discrepancies in tumor segmentation using
  diffusion models,''
\newblock {\em arXiv preprint arXiv:2402.07354}, 2024.

\bibitem{ren2023optimization}
T.~Ren, E.~Honey, H.~Rebala, A.~Sharma, A.~Chopra, and M.~Kurt,
\newblock ``An optimization framework for processing and transfer learning for
  the brain tumor segmentation,''
\newblock in {\em International Challenge on Cross-Modality Domain Adaptation
  for Medical Image Segmentation}, pp. 165--176, Springer, 2023.

\bibitem{diaconis1977spearman}
P.~Diaconis and R.~L. Graham,
\newblock ``Spearman's footrule as a measure of disarray,''
\newblock {\em Journal of the Royal Statistical Society Series B: Statistical
  Methodology}, vol. 39, no. 2, pp. 262--268, 1977.

\bibitem{mcknight2010mann}
P.~E. McKnight and J.~Najab,
\newblock ``Mann-whitney u test,''
\newblock {\em The Corsini encyclopedia of psychology}, pp. 1--1, 2010.

\bibitem{bland2015statistics}
J.~M. Bland and D.~G. Altman,
\newblock ``Statistics notes: bootstrap resampling methods,''
\newblock {\em bmj}, vol. 350, 2015.

\bibitem{wissler1905spearman}
C.~Wissler,
\newblock ``The spearman correlation formula,''
\newblock {\em Science}, vol. 22, no. 558, pp. 309--311, 1905.

\end{thebibliography}

\end{document}